\documentclass[10pt]{revtex4}
\usepackage{graphics, epsfig}

\begin{document}



\newcommand{\be}{\begin{equation}}

\newcommand{\ee}{\end{equation}}

\def\beqa{\begin{eqnarray}}

\def\eeqa{\end{eqnarray}}

\def\beq{\begin{equation}}

\def\eeq{\end{equation}}

\def\ad{\dot{a}}

\def\vol{\int d^4x\,\sqrt{-g}}

\def\grav{\frac{1}{16 \pi G}}

\def\half{\frac{1}{2}}

\def\gu{g^{\mu\nu}}

\def\gd{g_{\mu\nu}}

\def\umu{^{\mu}}

\def\unu{^{\nu}}

\def\dmu{_{\mu}}

\def\dnu{_{\nu}}

\def\umunu{^{\mu\nu}}

\def\dmunu{_{\mu\nu}}

\def\ua{^{\alpha}}

\def\ub{^{\beta}}

\def\da{_{\alpha}}

\def\db{_{\beta}}

\def\uamu{^{\alpha\mu}}

\def\uanu{^{\alpha\nu}}

\def\uab{^{\alpha\beta}}

\def\dab{_{\alpha\beta}}

\def\dabgd{_{\alpha\betaw\delta}}

\def\uabgd{^{\alpha\betaw\delta}}

\def\udeab{^{;\alpha\beta}}

\def\ddeab{_{;\alpha\beta}}

\def\ddemunu{_{;\mu\nu}}

\def\udemunu{^{;\mu\nu}}

\def\ddemu{_{;\mu}}  \def\udemu{^{;\mu}}

\def\ddenu{_{;\nu}}  \def\udenu{^{;\nu}}

\def\ddea{_{;\alpha}}  \def\udea{^{;\alpha}}

\def\ddeb{_{;\beta}}  \def\udeb{^{;\beta}}

\def\na{\nabla}

\def\naba{\nabla_{\alpha}}

\def\nabb{\nabla_{\beta}}

\def\pmu{\partial_{\mu}}

\def\pnu{\partial_{\nu}}

\def\pa{\partial}

\def\bib#1{$^{\ref{#1}}$}

\def\jmp{{\it J. Math. Phys.}\ }

\def\pr{{\it Phys. Rev.}\ }

\def\prl{{\it Phys. Rev. Lett.}\ }

\def\pl{{\it Phys. Lett.}\ }

\def\np{{\it Nucl. Phys.}\ }

\def\modpl{{\it Mod. Phys. Lett.}\ }

\def\ijmp{{\it Int. Journ. Mod. Phys.}\ }

\def\ijtp{{\it Int. Journ. Theor. Phys.}\ }

\def\cmp{{\it Commun. Math. Phys.}\ }

\def\cqg{{\it Class. Quantum Grav.}\ }

\def\ap{{\it Ann. Phys. (\mathcal{N}.Y.)}\ }

\def\spj{{\it Sov. Phys. JETP}\ }

\def\spjl{{\it Sov. Phys. JETP Lett.}\ }

\def\prs{{\it Proc. R. Soc.}\ }

\def\grg{{\it Gen. Relativ. Grav.}\ }

\def\nat{{\it Nature}\ }

\def\apj{{\it Ap. J.}\ }

\def\aa{{\it Astron. Astrophys.}\ }

\def\ncim{{\it Il Nuovo Cim.}\ }

\def\ptp{{\it Prog. Theor. Phys.}\ }

\def\aip{{\it Adv. Phys.}\ }

\def\jpamg{{\it J. Phys. A: Math. Gen.}\ }

\def\mnras{{\it Mon. Not. R. Ast. Soc.}\ }

\def\prep{{\it Phys. Rep.}\ }

\def\ncb{{\it Il Nuovo Cimento ``B''}}

\def\ssr{{\it Space Sci. Rev.}\ }

\def\pasp{{\it Pub. A. S. P.}\ }

\def\araa{{\it Ann. Rev. Astr. Ap.}\ }

\def\asr{{\it Adv. Space Res.}\ }

\def\rmp{{\it Rev. Mod. Phys.}\ }

\def\etal{{\it et al.}}

\def\ie{{\it i.e. }}

\def\eg{{\it e.g. }}

\def\jmp{{\it J. Math. Phys.}\ }

\def\lrr{{\it Liv. Rev. Rel.}\ }

\def\nat{{\it Nature}\ }

\def\apj{{\it Ap. J.}\ }

\def\mnras{{\it Mon. Not. R. Ast. Soc.}\ }

\def\araa{{\it Ann. Rev. Astr. Ap.}\ }

\def\rmp{{\it Rev. Mod. Phys.}\ }

\def\arns{{\it Ann. Rev. Nucl. Part. Sci.}\ }

\let\lam=\lambda  \let\Lam=\Lambda

\let\eps=\varepsilon

\let\gam=w

\let\alp=\alpha

\let\sig=\sigma

\def\vol{d^4x\,\sqrt{-g}}

\def\volbr{d^4x\,\sqrt{-\bar{g}}}

\def\grav{\frac{1}{16 \pi G}}

\def\half{\frac{1}{2}}

\def\gu{g^{\mu\nu}}

\def\gd{g_{\mu\nu}}

\def\gbru{{\bar{g}}^{\mu\nu}}

\def\gbrd{{\bar{g}}_{\mu\nu}}

\def\GAM{{W}}

\def\LAM{{\Lambda}}

\def\OME{{\Omega}}

\def\fpq{{\phi^{'2}}}

\def\fbr{{\bar{\phi}}}

\def\fbrd{{\dot{\bar{\phi}}}}

\def\fbrdd{{\ddot{\bar{\phi}}}}

\def\fbrp{{{\bar{\phi}}^{'}}}

\def\fbrpp{{{\bar{\phi}}^{''}}}

\def\fbrpq{{{\bar{\phi}}^{'2}}}

\def\VDEF{{V_{\phi}}}

\def\VBR{{\bar{V}}}

\def\VBRDEF{{\bar{V}_{\bar{\phi}}}}

\def\FDEF{{F_{\phi}}}

\def\FD{{\dot{F}}}

\def\FDD{{\ddot{F}}}

\def\tbr{{\bar{t}}}

\def\LBR{{\overline{L}}}

\def\EBR{{\overline{E}}}

\let\lb=\label

\renewcommand{\epsilon}{\varepsilon}

\let\no=\nonumber

\def\disp{\displaystyle}

\def\psiul{\overline\psi}

\def\pb{\not\!\partial}

\def\f{F(\phi)}

\def\fp{F'(\phi)}

\def\fpp{F''(\phi)}

\def\p{\phi}

\def\pv{\varphi}

\def\v{V(\phi)}

\def\vp{V'(\phi)}

\def\l{\cal L}
\def\case#1/#2{\frac{#1}{#2}}
\def\rf#1{(\ref{#1})}

\title{Cosmological dynamics of fourth order gravity}

\author{S Carloni$^\dag$,
P K S Dunsby$^{\dag\,\sharp}$ and A Troisi$^\diamond\natural$
}
\affiliation{$^\dag$Institut de Cincies de l'Espai (CSIC-IEEC)
Campus UAB -  Facultat de Ciencies
Torre C5  Parell,  2da Planta
E-08193 Bellaterra  (Barcelona)
Spain }
\affiliation{$^\sharp$\ Department of Mathematics and Applied\
Mathematics, University of Cape Town, South Africa and South African Astronomical Observatory,
Observatory Cape Town, South Africa .}
\affiliation{$^{\diamond}$ Dipartimento di Scienze Fisiche and
INFN, Sez. di Napoli, Universit\`a di Napoli "Federico II", Compl.
Univ. di Monte S. Angelo, Edificio G, Via Cinthia, I-80126 -
Napoli, Italy }

\affiliation{$^{\natural}$ Dipartimento di Ingegneria Meccanica,
Universit\`a di Salerno, via Ponte don Melillo , I- 84084 -
Fisciano (SA), Italy.}

\begin{abstract}
We discuss the dynamical system approach  applied to  Higher Order Theories of
Gravity. We show that once the theory of gravity has been specified,
the cosmological equations can be written as a first-order
autonomous system and we give several examples which illustrate the
utility of our method. We also discuss a number of results which have
appeared recently in the literature.
\end{abstract}
\date{\today}

 \maketitle

\section{Introduction}

Although there are many good reasons to consider General
Relativity (GR) as the best theory for the gravitational
interaction, in the last few decades the advent of precision
cosmology tests appears more and more to suggest  that this theory
may be incomplete. In fact, besides the well known problems of GR
in explaining the astrophysical phenomenology (i.e., the galactic
rotation curves and small scale structure formation),
cosmological data indicates an underlying cosmic acceleration of the
Universe which cannot be recast in the framework of GR without
resorting to additional exotic matter components. Several models have
been proposed  \cite{lambda-darksector} in order to address this problem and
currently the one which best fits all available observations (Supernovae Ia \cite{sneIa}, Cosmic Microwave
Background anisotropies \cite{cmbr}, Large Scale Structure
formation \cite{lss}, baryon oscillations \cite{baryon}, weak
lensing \cite{wl}), turns out to be the {\it Concordance Model}
in which a tiny cosmological constant is present \cite{astier} and
ordinary matter is dominated by a Cold Dark component.
However, given that  the $\Lambda$\,-\,CDM model is affected by
significant fine-tuning problems related to the vacuum energy scale,
it seems  desirable to investigate other viable theoretical schemes.

It is for these reasons that in recent years many attempts have been
made to generalize standard Einstein gravity. Among
these models the so-called Extended Theory of Gravitation (ETG)
and, in particular, {\em non-linear gravity
theories} or {\em higher-order theories of gravity} (HTG) have provided interesting 
results on both cosmological \cite{starobinsky80,ccct-ijmpd,review,cct-jcap,otha,collezfR}
and astrophysical \cite{cct-jcap,cct-mnras} scales. These models are based on
gravitational actions which are non-linear in the Ricci curvature
$R$ and$/$ or contain terms involving combinations of derivatives
of $R$ \cite{kerner,teyssandier,magnanoff}. The peculiarity of
these models is related to the fact that the gravitational field equations can be
recast in such a way that the higher order corrections provide an
energy\,-\,momentum tensor of geometrical origin describing an
``effective" source term on the right hand side of the standard Einstein field
equations  \cite{ccct-ijmpd,review}. In this scenario,
the cosmic acceleration can be shown to result from  such a new geometrical contribution
to the cosmic energy density budget, due to higher order corrections to
the Hilbert-Einstein Lagrangian.

Because the field equations resulting from HTG are extremely complicated,
the theory of dynamical systems provides a powerful scheme for
investigating the physical behaviour of such theories (see for example \cite{cdct:dynsys05,ScTnDynSys}). In fact, studying cosmologies
using the dynamical systems approach has the advantage of providing a
relatively simple method for obtaining exact solutions (even if these
only represent the asymptotic behavior) and  obtain a (qualitative) description of
the global dynamics of these models. Consequently, such an analysis allows for an efficient 
preliminary investigation of these theories, suggesting what kind of models deserve further investigation. 
Of particular importance are those theories that admit solutions that have an expansion history similar to the standard 
$\Lambda$CDM model and are therefore worth considering as background models for a description of the growth 
of structure in HTG \cite{SantePertSca}.

In this paper, using the Dynamical Systems Approach (DSA) approach
suggested by Collins and then by Ellis and Wainwright (see
\cite{ellisbook} for a wide class of cosmological models in the GR
context),  we develop a completely general scheme, which in principle allows one to analyze every fourth order gravity Lagrangian.
Our study generalizes
\cite{cdct:dynsys05}, which considered a generic power law function of the
Ricci scalar $f(R)\,=\,R^{n}$ and extends the general approach given in a recent paper
\cite{amendola:dynsys06}. Here a general analysis was obtained using a one\,-parameter
description of any $f(R)$ model, which unfortunately turns out to be somewhat misleading.

The aim of this  paper is to illustrate the general procedure for
obtaining a phase space analysis for any analytical $f(R)$
Lagrangian, which is regular enough to be well defined up to the
third  derivative in $R$. After a short preliminary discussion about
fourth order gravity, we will discuss this general procedure, giving
particular attention to clarifying the differences between our
approach and the one worked out in \cite{amendola:dynsys06}. In
order to illustrate these differences and the problems that exist in
\cite{amendola:dynsys06}, we will apply our method to two different
families of Lagrangian $R^p\exp{qR}$ and $R+\chi R^n$. The last
part of the paper is devoted to discussion and conclusions.
Unless otherwise specified, we will use natural units
($\hbar=c=k_{B}=8\pi G=1$) and the $(+,-,-,-)$ signature.

\section{Fourth Order Gravity Models}
If one relaxes the assumption of linearity of the gravitational action the most general  fourth order Lagrangian in an homogeneous and isotropic spacetime can be written as\,:
\begin{equation}\label{lagr f(R)}
L=\sqrt{-g}\left[ f(R)+{\cal L}_{M}\right]\;.
\end{equation}
By varying equation (\ref{lagr f(R)}), we obtain the fourth order
field equations \beq\label{3.7.2}
f'(R)R_{\mu\nu}-\frac{1}{2}f(R)g_{\mu\nu}=f'(R)\udeab\left(g_{\alpha\mu}g_{\beta\nu}-
g\dab\gd\right)+\tilde{T}^{M}_{\mu\nu}\,, \eeq where
$\displaystyle{\tilde{T}^{M}_{\mu\nu}=\frac{2}{\sqrt{-g}}\frac{\delta
(\sqrt{-g}L_{M})}{\delta g_{\mu\nu}}}$ and the prime denotes the
derivative with respect to $R$. Standard Einstein equations are
immediately recovered if $f(R)=R$. When $f'(R)\neq0$ the equation
(\ref{3.7.2}) can be recast in the form
\begin{equation}\label{73-curv3}
G_{\mu\nu}=R_{\mu\nu}-\frac{1}{2}g_{\mu\nu}R=T^{TOT}_{\mu\nu}=T^{R}_{\mu\nu}+T^{M}_{\mu\nu}\,,
\end{equation}
where
\begin{equation}
\label{74-curv4} 
T^{R}_{\mu\nu}=\frac{1}{f'(R)}\left\{\frac{1}{2}g_{\mu\nu}\left[f(R)-Rf'(R)\right]+
f'(R)^{;\alpha\beta}(g_{\alpha\mu}g_{\beta\nu}-g_{\alpha\beta}g_{\mu\nu})
\right\}\;,
\end{equation}
represent the stress energy tensor of an effective fluid sometimes referred to as
the  ``curvature fluid" and
\begin{equation}
\label{75-curv5}
T^{M}_{\mu\nu}=\frac{1}{f'(R)}\tilde{T}^{M}_{\mu\nu}\;,
\end{equation}
represents an effective stress-energy tensor associated with
standard matter.

The conservation properties of these effective fluids are given in \cite{SantePertSca,DynSysGen}  but it is important to stress that even if the {\em effective tensor} associated with the matter is not conserved, standard
matter still follows the usual conservation equations
$\tilde{T}_{\mu\nu}^{M;\nu}=0$.

Let us now  consider the Friedmann-Lema\^{\i}tre-Robertson-Walker
(FLRW) metric:
\begin{equation}\label{frw}
 ds^2 = dt^2 - a^2(t)\left[ {dr^2 \over 1-kr^2} + r^2 (d\theta^2 +
\sin^2\theta d\phi^2)\right]\;.
\end{equation}
For this metric the action the field equations (\ref{74-curv4})
reduce to
\begin{eqnarray}\label{E12}
& H^{2}+\frac{k}{a^2} =
\frac{1}{3f'}\left\{\frac{1}{2}\left[f'R-f\right]-3H\dot{f'}+\mu_{{m}}\right\}\,,\\
&2\dot{H}+H^{2}+\frac{k}{a^2} =
-\frac{1}{f'}\left\{\frac{1}{2}\left[f'R-f\right]+\ddot{f'}-3H\dot{f'}+\,p_{{m}}\right\}\,,
\end{eqnarray}
and
\begin{equation}
R\,=\,-6\left(2H^{2}+\dot{H}+\frac{k}{a^2}\right)\,,\label{R}
\end{equation}
where $H\equiv\dot{a}/a$, $f'\equiv\frac{d f(R)}{d R}$ and the
``dot" is the derivative with respect to $t$. The system \rf{E12}
is closed by the Bianchi identity for
$\tilde{T}^{M}_{\mu\nu}$:
\begin{equation}
\dot{\mu}_{m}+3H(\mu_{ m}+p_{m})=0\;,\label{E3}
\end{equation}
which corresponds to the energy conservation equation for standard matter.

\section{The dynamical system approach in fourth order gravity
theories}\label{dynsysSect}

Following early attempts (see for example \cite{dys4OrGrav}), the first
extensive analysis of cosmologies based on fourth order gravity
theory using the DSA as defined in
\cite{ellisbook} was given in \cite{cdct:dynsys05}. Here
the phase space of the power law model $f(R)\,=\,\chi R^n$ was
investigated in great detail,  exact solutions were found and their stability determined.
Following this, several authors have applied a similar approach to other types of
Lagrangians \cite{barrow:dynsys}, and very recently this scheme was
generalized in \cite{amendola:dynsys06}. 

In this paper we give a self consistent general technique that allows us to perform a
dynamical system analysis of any analytic fourth order theory of gravity in the case of the FLRW spacetime.

The first step in the implementation of the DSA is the definition of the variables. Following
\cite{cdct:dynsys05}, we introduce the general dimensionless
variables \,:
\begin{eqnarray}\label{phi2:var}
x = \frac{\dot{f'}}{f' H}, \qquad y = \frac{R}{6 H^2},  \qquad z = \frac{f}{6 f' H^2},  \qquad 
\Omega = \frac{\mu_m}{3 f' H^2},  \qquad K =\frac{k}{a^2 H^2}\;,
\end{eqnarray}
where $\mu_m$ represents the energy density of a perfect fluid
that might be present in the model.

The cosmological equations (\ref{E12}) are equivalent to the
autonomous system\,:
\begin{eqnarray}\label{dynsysK}
\frac{dx}{d N} &=& \varepsilon\,(2 K+2 z-x^2+(K+y+1) x)+\Omega\varepsilon\, (-3 w -1)+2, \\
\frac{dy}{d N} &=& y\varepsilon\,(2 y+2 K+x \Upsilon+4) , \\
\frac{dz}{d N} &=& z\varepsilon\,(2 K-x+2 y+4) + \varepsilon\, x y\Upsilon, \\
\frac{d\Omega}{d N} &=& \Omega\varepsilon\, (2 K-x+2 y-3 w+1),\\
\frac{d K}{d N} &=&2K\varepsilon\,(K+y+1) ,
\end{eqnarray}
where $N=|\ln a|$ is the
logarithmic time and $\varepsilon=|H|/H$. In addition, we have the constraint equation
\begin{equation}
    1=-K-x-y+z+\Omega\,,
\end{equation}
which can be used to reduce the dimension of the system. If one
chooses to eliminate $K$, the variable associated with the spatial
curvature, we obtain
\begin{eqnarray}\label{DynsysNoK}
\frac{dx}{d N} &=&\varepsilon\,(4 z-2 x^2+(z-2) x-2 y)+\Omega\varepsilon\, (x-3 w +1), \nonumber \\
\frac{dy}{d N} &=&y \varepsilon\,[2 \Omega+2 (z+1)+x(\Upsilon-2)],  \\
\frac{dz}{d N} &=&z\varepsilon\,(2 z+2 \Omega-3 x+2) z+x\varepsilon\, y \Upsilon, \nonumber \\
\frac{d\Omega}{d N} &=& \Omega \,\varepsilon\,(2 \Omega -3 x+2 z-3 w -1), \nonumber \\
K &= &  z + \Omega - x- y - 1 \;. \nonumber
\end{eqnarray}
The quantity $\Upsilon$ 
is defined, in analogy with \cite{amendola:dynsys06}, as
\begin{equation}\label{q}
\Upsilon\,\equiv
\,\frac{f'}{Rf''}\,.
\end{equation}
The expression of $\Upsilon$ in terms of the dynamical variables
is the key to closing the system \rf{dynsysNoK} and allows one to perform
the analysis of the phase space. The crucial aspect to note here is
that $\Upsilon$ is a function of $R$ only, so the problem of
obtaining $\Upsilon=\Upsilon(x,y,z,\Omega)$ is reduced to
the problem of writing $R=R(x,y,z,\Omega)$. This can be achieved by
noting that the quantity
\begin{equation}\label{Pre-r}
r\,\equiv
\,-\frac{Rf'}{f}\,,
\end{equation}
is a function  of $R$ only and can be written as
\begin{equation}\label{r}
   r =-\,\frac{y}{z}\,.
\end{equation}
Solving the above equation for $R$ allows one to write $R$ in terms of
$y$ and $z$ and close the system \rf{DynsysNoK}.

In this way, once a Lagrangian has been chosen,  we can in principle write
the dynamical system associated with it using \rf{DynsysNoK},
substituting into it the appropriate form of $\Upsilon=\Upsilon(y,z)$.
This procedure does however require particular attention. For example, there are
forms of the function $f$ for which the inversion of \rf{r} is highly non trivial (e.g.,
$f(R)=\cosh(R)$). In addition, the function $\Upsilon$ could
have a non-trivial domain, admit divergences or may not be in the
class $C^{1}$, which makes the analysis of the phase space a
very delicate problem. Finally, the number $m$ of equations of
\rf{DynsysNoK} is always $m\geq3$ and this implies that fourth order
gravity models can admit  chaotic behaviour. While this is not surprising, it
makes the deduction of the non--local properties of the phase space
a very difficult task.

The solutions associated with the fixed points can be found by
substituting the coordinates of the fixed points into the system
\begin{eqnarray}\label{solsystem}
\dot{H} &=&\alpha H^{2}\;, \qquad \alpha=-1 - \Omega_i + x_i - z_i\,,\label{solsystem1}\\ 
\dot{\mu}_m &=& -\frac{3(1+w)}{\alpha\; t}\mu_m \label{solsystem2}\,,
\end{eqnarray}
where the subscript ``$i$" stands for the value of a generic quantity in a fixed point. 
This means that  for $\alpha\neq 0$ the general solutions  can be written as 
\begin{eqnarray}\label{solsystemalp}
a &=&a_{0}(t-t_{0})^{1/\alpha }\;,  \\
\mu_m &=& a_{0}(t-t_{0})^{-\frac{3(1+w)}{\alpha}} \,.
\end{eqnarray}
The  expression above  gives the solution for the scale factor and the evolution 
of the energy density for every fixed point in which  
$\alpha\neq 0$. When $\alpha=0$ the \rf{solsystem1} reduces to
$\dot{H}=0$ which correspond to either a static or a de Sitter
solution.

The solutions obtained in this way have to be considered particular
solutions of the cosmological equations  which are found by using a specific ansatz (i.e. the fixed point condition  \cite{DynSysPert}). For this reason it is important 
to stress that only direct substitution of the results derived from  
this approach in the cosmological equations can ensure that the solution is physical (i.e. it satisfies the cosmological
equations \rf{E12} ). This check is also useful for understanding
the nature of the solutions themselves e.g., to calculate the
value of the integration constant(s).

Also, the fact that different fixed points correspond to the
same solutions is due to the fact that at the fixed points the
different terms in the equation combine in such a way to obtain the
same evolution of  the scale factor. This means that although two
solutions are the same in terms of time dependence, the physical mechanism that realizes them can be different

One difference betwen our approach and the one in  \cite{amendola:dynsys06} 
is that we consider  a non-zero spatial curvature $k$. The choice of including a non-zero spatial curvature $k$ has been made with the aim of obtaining a
completely general analysis of a fourth order cosmology from the dynamical systems point of view. In addition,
since most of the observational values for the cosmological parameters are heavily model dependent, we chose to
limit as much as possible the introduction of priors in the analysis. However, as we write in the footnote in
section 3, the limit of flat spacelike sections ($K\rightarrow 0$) can be obtained in a straightforward way for
our examples. In fact, each fixed point is associated with  a specific value of the variable $K$ (i.e. a value for
$k$) and the stability of these points is independent of the value of $K$. As matter of fact  in order to
consider fixed points living on the hypersurface $K = 0$, one has just to exclude the fixed points associated with
$K\neq 0$. In addition to that, looking at the dynamical equations one realizes that $K=0$ is an invariant
submanifold, i.e., an orbit with initial condition $K=0$ will not escape the subspace $K=0$ and orbits with initial
condition $K\neq0$ can approach the hyperplane $K=0$ only asymptotically. As a consequence, one does not need to
have any other information on the rest of the phase space to characterize the evolution of the orbits in the
submanifold $K=0$. 
The authors of \cite{amendola:dynsys06} proposed that the
function $m(r)=\Upsilon(r)^{-1}$ could be used as a parameter associated with the choice of
$f(R)$, thus obtaining a  ``one parameter approach"  to the dynamical systems
analysis of $f(R)$ gravity. Unfortunately their method has several problems that
lead to incorrect results. These problems can be avoided only if one considers the framework  presented above.

Let us look at this issue in more detail \footnote{It is important to
note that in \cite{amendola:dynsys06} the signature is not the same
of the one used here (e.g -,+,+,+ instead of +,-,-,-) and the
definition of the variables are slightly different. The
transformation from one variable to another is as follows:  $$
x\rightarrow -x_1,\quad y\rightarrow -x_3, \quad z\rightarrow x_2, 
\quad K\rightarrow 0,\quad w\rightarrow 0.$$ However, as expected,
this does not affect our conclusions.}. In \cite{amendola:dynsys06}
the system equivalent to \rf{dynsysNoK} is associated with the relation
\begin{equation}
\frac{dr}{dN}=r(1+m(r)+r)\frac{\dot{R}}{HR}\,,\label{eq:critline}
\end{equation}
which is clearly  a combination of the equations for $z$ and $y$. In
order to ensure that the variable $r$ and consequently the parameter
$m$ is constant they require the RHS of the above equation to be
zero. Their solution to this problem is the condition $1+m(r)+r=0$,
which is an equation for $r$ when the function $m(r)$ has been
substituted for and is also the bases of their method of analysis.

The problem here is that this equation has not been fully expressed in terms
of the  dynamical system variables. In fact, one can rewrite
(\ref{eq:critline}) in the form\,:
\begin{equation}
\frac{dr}{dN}=\frac{r(1+m(r)+r)}{m(r)}x\;,\label{eq:critline2}
\end{equation}
which means that the condition $\displaystyle\frac{dr}{dN}=0$ in
fact corresponds to
\begin{equation}\label{eq:critcond2}
   \frac{ r(1+m(r)+r)}{m(r)}x=0/,,
\end{equation}
rather than $1+m(r)+r=0$. Equation \rf{eq:critcond2} has a solution if
\begin{eqnarray}
&&x=0, \\
&&r=0, \\
&&\frac{(1+m(r)+r)}{m(r)}\,=\,0\,,
\end{eqnarray}
and this leads to solutions for $r$ which are in general different from the values of $r$ obtained from $1+m(r)+r=0$. This inconsistency has major
consequences for the rest of the analysis in \cite{amendola:dynsys06}, leading to changes in the number of fixed points as well as their
stability (see below for details).

In fact, a more careful analysis reveals that for some of the fixed
points (e.g. $P_1,...P_4$) the values of $r$ obtained from the
relation $r=-y/z$ either cannot be determined unambiguously or do
not solve the condition $1+m(r)+r=0$, which is claimed
to come from (\ref{eq:critline}) in \cite{amendola:dynsys06}.

This is a clear indication that the approach used in
\cite{amendola:dynsys06} is both incomplete and leads to
wrong conclusions. It is also interesting to stress that if one
substitutes the expression for $m$ in terms of the dynamical system
variables in (26-29) of \cite{amendola:dynsys06}, the results match
the one obtained in our formalism. This implies that the reason
the method described in \cite{amendola:dynsys06} fails has its roots
in the attempt to describe the phase space of a whole class of fourth order
theories of gravity with only one parameter.

In the following we will present a number of examples of $f(R)$ theories
that can be analyzed with this method and we compare the results
obtained with those given in \cite{amendola:dynsys06}.

\section{Examples of $f(R)$\,-\,Lagrangians}

In this section  we will show, with the help of some examples, how the DSA 
developed above can be applied. In particular we
will consider the cases $f(R)\,=\,R^p\exp(qR)$ and $f(R)\,=\,R+\chi
R^n$. Since the aim of the paper is to provide only the general
setting with which to develop the dynamical system approach in the
framework of fourth order gravity, we will not give a detailed
analysis of these models. Istead,  we will limit ourselves to
the finite fixed points, their stability and the solutions
associated with them. A comparison with the results of
\cite{amendola:dynsys06} will also be presented.

\subsection{The $f(R)\,=\,R^p\exp(q R)$ case}

Let us consider the Lagrangian $f(R)\,=\,R^p\exp(q R)$. As explained
in the previous section, the dynamical system equations for this
Lagrangian can be obtained by calculating the form of the
parameter $\Upsilon$. We have
\begin{equation}
\Upsilon(y,z)\,= \,\frac{y\;z}{y^2 -p \;z^2}\,.
\end{equation}
Substituting this function into \rf{dynsysNoK} we obtain
\begin{eqnarray}\label{dynsysNoKExp}
\frac{dx}{d N} &=& \varepsilon\,[4 z-2 x^2+(z-2) x-2 y]+\Omega \varepsilon\,(x-3 w +1), \\
\frac{dy}{d N} &=&y\varepsilon\, \left[2 \Omega+2 z+2+\frac{x\;z}{y^2 -p\;z^2}-2x\right], \\
\frac{dz}{d N} &=& z\varepsilon\,\left[2 z+2 \Omega-3 x+2+ \frac{x\;y}{y^2 -p\; z^2}\right], \\
\frac{d\Omega}{d N} &=& \Omega \,\varepsilon\,(2 \Omega -3 x+2 z-3 w -1),\\
 K &=&   z + \Omega - x- y - 1 \,.
\end{eqnarray}
The most striking feature of this system is the fact that two of the
equations have a singularity in the hypersurface $y^2 =p\; z^2$.
This, together with the existence of the invariant submanifolds $y=0$
and $z=0$ heavily constrains the dynamics of the system. In
particular, it implies that no global attractor is present, thus no
general conclusion can be made on the behavior of the orbits without
first providing information about the initial conditions. The finite fixed
points can be obtained by setting the LHS of \rf{dynsysNoKExp} to zero
and solving for $(x,y,z,\Omega)$, the results are shown in Table
\ref{tabRexpfixp}.

The solutions corresponding to these fixed points can be obtained by
substituting the coordinates into  the system \rf{solsystem} and are
shown in Table \ref{tabRexpSol} \footnote{Note that even if the
parameter $q$ is not present in the dynamical equations it appears
in the solutions because we have calculated the integration
constants via direct substitution in the cosmological equations.}.
The stability of the finite fixed points can be found using the
Hartman-Grobman theorem \cite{HG}. The results are shown in Table
\ref{tabRexpStab}. Note that some of the eigenvalues diverge for
$p=0,1$. This happens because in the operations involved in the derivation of the stability terms 
$p-1$ and/or $p$ appear in the denominators. However this is not a real pathology of the method but rather a consequence 
of the fact that for these two values of the  parameter the cosmological equations assume a special form. 
In fact it is easy to prove that if one starts the calculations using   
these critical values of $p$  one ends up with eigenvalues that present no divergence \cite{DynSysGen}.

Let us now compare our results with the ones in
\cite{amendola:dynsys06}. 
The number of fixed points obtained for this Lagrangian, when $K\,=\,0$, matches the ones obtained in \cite{amendola:dynsys06}. This result can be
explained by the fact that the solutions of the constraint equation
for $m$ (\ref{eq:critline}) coincide with the ones coming from the
correct constraint equation (\ref{eq:critline2}) (the matching
between the two systems can be obtained setting $w \,=\,0$ in Table
\ref{tabRexpfixp}). However, when one calculates the stability of
these points our results are strikingly different to those
presented in \cite{amendola:dynsys06}. For example, in our general
formalism it turns out that the fixed point $\mathcal{N}$ (corresponding to $P_5$ of \cite{amendola:dynsys06}) is a
saddle for any value of the parameter $p$ and, as consequence, it
can represent only a transient phase in the evolution of this class
of models. Instead, in \cite{amendola:dynsys06} the authors find
that this point can be  stable (not necessarily always a spiral) and argue that this fact prevents the
existence of cosmic histories in which a decelerated expansion is
followed by an accelerated one. From this they also conclude that an
entire subclass of these models ($m\,=\,m(p)>0$) can be ruled out.
Our results show clearly that this is not the case. Another example
is the point $\mathcal{\mathcal{M}}$ corresponding to $P_6$ of \cite{amendola:dynsys06} . In
\cite{amendola:dynsys06} the authors find that this point  can be
stable or a saddle as we do, but  the intervals of values of the parameters for which this happens are different (see Table \ref{tabRexpStab}). As explained above, the reason behind these differences is the fact that the
method used in \cite{amendola:dynsys06} leads to incorrect results
when, like in this case, there is no unambiguous way of
determining the parameter $r\,=\,-y/z$ from the coordinates of the
fixed points. Consequently the conclusions in
\cite{amendola:dynsys06} relating to the properties of these points
are incorrect and have no physical meaning.

\begin{table}[h]
\caption{Fixed points of $R^{p}\exp(q R)$.The superscript ``*" represents a point
corresponding to a double solution.} \label{tabRexpfixp}
\begin{tabular}{llrl} \hline\hline
Point & Coordinates $(x,y,z,\Omega)$ & $K$ &
\\ \hline
$\mathcal{A}$ & $\left(0, 0, 0, 0\right)$ & $-1$ &   \\
$\mathcal{B}$ & $\left(-1, 0, 0, 0\right)$
& $0$ &  \\
$\mathcal{C}$ &
$\left(-1-3w,0,0,-1-3w\right)$ & $-1$ & $$\\
$\mathcal{D}$ &
$\left(1-3w,0,0,2-3w\right)$ & $0$ & $$\\
$\mathcal{E}$ & $\left(2,0,2,0\right)$ & $-1$ &
$$\\
$\mathcal{F}^{*}$ & $\left(1, -2, 0, 0\right)$& $0$&
 \\
$\mathcal{G}$ & $\left(0,-2,-1,0\right)$ & $0$ &
$$
 \\
$\mathcal{H}$ & $\left(4,0,5,0\right)$ & $0$ &
$$\\
$\mathcal{I}^{*}$ & $\left(-3(1+w),-2,0,-4-3w\right)$ & $0$ & $$ \\ 
$\mathcal{L}$ & $\left(2-2p,2p(1-p),2-2p,0\right)$ & $2p(p-1)-1$ &
$$
 \\

 $\mathcal{M}$ & $\left(\frac{4-2p}{1-2p},\frac{(5-4 p) p}{2 p^2-3
   p+1},\frac{5-4 p}{(p-1) (2 p-1)},0\right)$ & $0$& $ $\\
$\mathcal{\mathcal{N}}$&
$\left(\frac{-3(1+w)(p-1)}{p},\frac{3(1+w)-4p}{2p},\frac{-4 p+3 w
+3}{2 p^2},\frac{p (9 w -2 p (3 w +4)+13)-3 (w +1)}{2 p^2}\right)$ &
$0$ & $$  \\ \hline\hline
 \end{tabular}
\end{table}

\begin{table}[h]
\caption{Solutions associated with the fixed points of $R^{p}\exp(q
R)$. The solutions are physical only in the intervals of $p$
mentioned in the last column.} \label{tabRexpSol}
\begin{tabular}{lccc} \hline\hline
Point & Scale Factor & Energy Density& Physical
\\ \hline
$\mathcal{A}$ & $a(t)=  \left(t-t_0\right)$ & $0$ &  $p\geq1$\\
$\mathcal{B}$ & $a(t)= a_0 \left(t-t_0\right)^{1/2}$ & $0$&  $p\geq2$\\
$\mathcal{C}$ &$a(t)= \left(t-t_0\right)$& $0$&$p\geq1$ \\
$\mathcal{D}$ &$a(t)= a_0 \left(t-t_0\right)^{1/2}$& $0$&  $p\geq2$\\
$\mathcal{E}$ & $a(t)=  (t- t_0)$ & $0$ & $p\geq1$\\
$\mathcal{F}^{*}$ & $\left\{
                      \begin{array}{l}
                        a(t)= a_0,  \\
                        a(t)=a_0\exp\left[\pm\frac{\sqrt{2-3p}}{6\sqrt{q}}(t-t_{0})\right],
                        \end{array}\right.$&0&$\begin{array}{c}
                                     p\geq0\\
                                     p<\frac{2}{3}, q>0 \vee p>\frac{2}{3}, q<0
                                   \end{array}$\\
$\mathcal{G}$ & $\left\{
                      \begin{array}{l}
                        a(t)= a_0,  \\
                        a(t)=a_0\exp\left[\pm\frac{\sqrt{2-3p}}{6\sqrt{q}}(t-t_{0})\right],
                        \end{array}\right.$& 0&$\begin{array}{c}
                                     p\geq0 \\
                                     p<\frac{2}{3}, q>0 \vee p>\frac{2}{3}, q<0
                                   \end{array}$\\
$\mathcal{H}$ & $a(t)= a_0 \left(t-t_0\right)^{1/2}$ & 0 &  $p\geq2$\\
$\mathcal{I}^{*}$ & $\left\{
                      \begin{array}{l}
                        a(t)= a_0,  \\
                        a(t)=a_0\exp\left[\pm\frac{\sqrt{2-3p}}{6\sqrt{q}}(t-t_{0})\right],
                        \end{array}\right.$&0&$\begin{array}{c}
                                     p\geq0 \\
                                     p<\frac{2}{3}, q>0 \vee p>\frac{2}{3}, q<0
                                   \end{array}$\\
$\mathcal{L}$ & $a(t)= \left(t-t_0\right)\sqrt{1-2p(p-1)} $ & $0$&$1\leq p\leq \frac{1}{2}+\frac{\sqrt{3}}{2}$\\$\mathcal{M}$ & $a(t)=a_0\left(t-t_0\right)^{\frac{2 p^2-3
p+1}{2-p}}$  & $\mu_m=\mu_{m\,0} t^{\frac{3 \left(2 p^2-3 p+1\right) (w +1)}{p-2}}$&$p=\frac{1}{2},1,\frac{5}{4}$\\
$\mathcal{\mathcal{N}}$& $a(t)= a_0\left(t-t_0\right)^{\frac{2 p}{3
(w
+1)}}$  & $\mu_m=\mu_{m\,0} (t-t_{0})^{-2 p} $ & $p=\frac{3 (w +1)}{4}\;\;\; (\mu_{m\,0}=0)$\\
\hline\hline
 \end{tabular}
\end{table}
\begin{table}[h] \centering \caption{The stability  associated
with the fixed points in the model $R^{p}\exp(qR)$. With the index
$^+$ we have indicated the attractive nature of the spiral points.
\label{tabRexpStab}}
\begin{tabular}{ll}
\hline\hline Point &  Stability \\ \hline
& \\
$\mathcal{A}$ &   saddle   \\
$\mathcal{B}$ & $\left\{\begin{array}{cc}
                  \mbox{repellor} & 0<w<2/3 \\
                   \mbox{saddle} &  \mbox{otherwise}
                \end{array}\right.$ \\
$\mathcal{C}$ & saddle  \\
$\mathcal{D}$ &  $\left\{\begin{array}{cc}
                  \mbox{repellor} & 2/3<w<1 \\
                   \mbox{saddle} &  \mbox{otherwise}
                \end{array}\right.$ \\
$\mathcal{E}$ & saddle  \\
$\mathcal{F}$ & saddle  \\
$\mathcal{G}$ & $\left\{\begin{array}{cc}
                   \mbox{attractor} & 0<w<1\cup2<p\leq\frac{68}{25} \\
                   \mbox{spiral}^+ & 0\leq w\leq1\cup\frac{68}{25}<p<4 \\
                    \mbox{saddle} & \mbox{otherwise}
                 \end{array}\right.$\\
$\mathcal{H}$ & saddle \\
$\mathcal{I}$& non hyperbolic \\
$\mathcal{L}$ & $\left\{\begin{array}{cc}
                   \mbox{attractor} &  
                   \frac{1}{2}-\frac{\sqrt{3}}{2}<p\leq{0} \vee \frac{4}{3}\leq{p}<\frac{1}{2}+\frac{\sqrt{3}}{2} \\
                   \mbox{spiral}^+ & 
                   {0}<p<\frac{4}{3}\\
                    \mbox{saddle} & \mbox{otherwise}
                 \end{array}\right.$\\
$\mathcal{M}$ &
$\left\{\begin{array}{cc}
                  \mbox{attractor} &  
                  p<\frac{1}{2}(1-\sqrt{3}) \vee \frac{1}{2}(1+\sqrt{3})<p<2 \\
                   \mbox{saddle} &  \mbox{otherwise}
                \end{array}\right.$\\
$\mathcal{\mathcal{N}}$ & saddle\\
\hline\hline
\end{tabular}
\end{table}

\subsection{The case $f(R)\,=R+\chi R^n$}
Let us discuss now the case of a Lagrangian corresponding to a power
law correction of the Hilbert\,-\,Einstein gravity Lagrangian
$f(R)\,=\,R+\chi R^n$. In this case, the characteristic function
$\Upsilon(y,z)$ reads\,:
\begin{equation}\label{m-R+Rngen}
\Upsilon(y,z)\,=\,\frac{y}{n(z-y)}\,,
\end{equation}
and substituting this relation into the system of equations
(\ref{dynsysNoK}) one obtains
\begin{eqnarray}\label{dynsysNoK}
\frac{dx}{d N} &=& -2 x^2+(z-2) x-2 y+4 z+\Omega (x-3 w +1), \\
\frac{dy}{d N} &=& y \varepsilon\,[2 \Omega+2 (z+1)+\frac{x\,y}{n(z-y)}-2x], \\
\frac{dz}{d N} &=&7 z\varepsilon\,(2 z+2 \Omega-3 x+2) +\varepsilon\, \frac{x\;y^{2}}{n(z-y)}, \\
\frac{d\Omega}{d N} &=& \Omega \,\varepsilon\,(2 \Omega -3 x+2 z-3 w -1),\\
K &=&   z + \Omega - x- y - 1 \,.
\end{eqnarray}
As in the case of $f(R)=R^{p}\exp(q R)$, the system is divergent on
a hypersurface (this time $y=z$) but it admits only one invariant
submanifold, namely $y=0$. This, again, implies that no global
attractor is present and no general conclusion can be made on the
behavior of the orbits without giving information about the initial
conditions. The finite fixed points,  their
stability and the solutions corresponding to them are summarized in
Tables \ref{fixpointR+Rn}, \ref{tabR+RnStab} and
\ref{tabRRnSol}.

As before our results are different from those given in
\cite{amendola:dynsys06}. First of all, our set of fixed points
do not  coincide with the ones presented in
\cite{amendola:dynsys06}. In particular, in our analysis there is no
fixed point corresponding to $P_{5a}$.  Again, the reason for this
difference is to be found in the constraint equation
(\ref{eq:critline}), which in this case gives the incorrect set of
solutions and therefore affects the set of fixed points. In fact, if one
substitutes the expression for $m(r)$ of \cite{amendola:dynsys06} in terms of the
coordinates in equations (34)-(39), it is easy to verify that  two of these
equations diverge at this point.

The differences between the results in our approach and the one
presented in \cite{amendola:dynsys06} are even more evident when the
stability analysis is considered.  For example, the point
$\mathcal{E}$, corresponding to $P_1$, is always a saddle, except
into the region $ 0< n < 2$ when it is attractive. This
behavior is recovered in \cite{amendola:dynsys06}  only for $-2<n<-41/25$. Also, points $\mathcal{G}$
(corresponding to $P_4$ of \cite{amendola:dynsys06}) and $\mathcal{D}$ (corresponding to $P_3$ of \cite{amendola:dynsys06}), which in our approach are always
saddles in the dust case, are always
repellers in \cite{amendola:dynsys06}. Finally, also the stability of
$\mathcal{I}$ corresponding to $P_6$ appears to be different from
the one presented in \cite{amendola:dynsys06}.
\begin{table}[h]
\caption{Coordinate of the finite fixed points for $R+\chi
R^{n}$ gravity.\label{fixpointR+Rn}}
\begin{tabular}{llrl} \hline\hline
Point & Coordinates $(x,y,z,\Omega)$ & $K$ &
\\ \hline
$\mathcal{A}$ & $\left(0, 0, 0, 0\right)$ & $-1$ &   \\
$\mathcal{B}$ & $\left(-1, 0, 0, 0\right)$
& $0$ &  \\
$\mathcal{C}$ &
$\left(-1-3w,0,0,-1-3w\right)$ & $-1$ & $$\\
$\mathcal{D}$ & $\left(1-3w,0,0,2-3w\right)$ & $0$ &
$$\\
$\mathcal{E}$ & $\left(0, -2, -1, 0\right)$& $0$&
 \\
$\mathcal{F}$ & $\left(2,0,2,0\right)$ & $-1$ & $$
 \\
 $\mathcal{G}$ & $\left(4,0,5,0\right)$ & $0$ & $$ \\
 $\mathcal{H}$ & $\left(2(1-n),2n(n-1),2(1-n),0\right)$ & $2n(n-1)-1$ & $$\\
$\mathcal{I}$ & $\left(\frac{2 (n-2)}{2 n-1},\frac{(5-4 n) n}{2
n^2-3 n+1},\frac{5-4 n}{2 n^2-3 n+1},0\right)$ & $0$ & $ $ \\
$\mathcal{L}$& $\left(-\frac{3 (n-1) (w +1)}{n},\frac{-4 n+3 w +3}{2
n},\frac{-4 n+3w +3}{2 n^2},\frac{-2 (3 w +4) n^2+(9 w +13) n-3 (w
+1)}{2 n^2}\right)$ & $0$& $ $ \\ \hline\hline
 \end{tabular}
\end{table}
\begin{table}[ht] \centering \caption{The stability of  the fixed points in the model $R+\chi R^n$. The quantities
$B_i$ related to the fixed point $\mathcal{L}$, represent some non
fractional numerical values ($B_1\approx1.220$, $B_1\approx1.224$,
$B_3\approx1.470$). \label{tabR+RnStab}}
\begin{tabular}{ll}
\hline\hline Point &  Stability \\ \hline
& \\
$\mathcal{A}$ &
saddle  \\
$\mathcal{B}$ & $\left\{\begin{array}{cc}
                  \mbox{repellor} &  0<w<2/3\\
                  \mbox{saddle} & \mbox{otherwise}
                \end{array}\right.$\\
$\mathcal{C}$ & saddle  \\
$\mathcal{D}$ & $\left\{\begin{array}{cc}
                  \mbox{repellor} &  2/3<w<1\\
                  \mbox{saddle} & \mbox{otherwise}
                \end{array}\right.$\\
$\mathcal{E}$ & $\left\{\begin{array}{cc}
                  \mbox{attractor} &  \frac{32}{25}\le n < 2\\
                  \mbox{spiral}^+ &  0 < n < \frac{32}{25}\\
                  \mbox{saddle} & \mbox{otherwise}
                \end{array}\right.$\\
$\mathcal{F}$ & saddle  \\
$\mathcal{G}$ & saddle \\
$\mathcal{H}$ & $\left\{\begin{array}{cc}
                  \mbox{attractor} & \frac{1}{2}(1-\sqrt{3})<n\le 0\\
                  \mbox{spiral}^+ &  0<n<1\\
                  \mbox{saddle} & \mbox{otherwise}
                \end{array}\right.$\\
$\mathcal{I}$ & $\left\{\begin{array}{lc}
 \mbox{attractor}   & n<\frac{1}{2}(1-\sqrt{3})\cup n>2 ,  \\
                                      \mbox{repeller}  & \left\{\begin{array}{cc}
                                                                  & 1<n<\frac{5}{4},  (w=0,1/3),\\
                                                                  &1<n<\frac{1}{14}(11+\sqrt{37}),  (w=1) 
                                                                   \end{array}\right. \\
                                      \mbox{saddle}      &  \mbox{otherwise},\\
                \end{array}\right.$\\
$\mathcal{L}$ &  $\left\{\begin{array}{lc}
                  w=0,1/3   & \mbox{saddle}, \\
                  w=1   & \left\{\begin{array}{cc}
                  \mbox{repellor} &  B_1<n\le B_2\cup B_3<n<\frac{3}{2} ,\\
                  \mbox{saddle} & \mbox{otherwise}
                \end{array}\right.
                \end{array}\right.$\\\hline\hline
\end{tabular}
\end{table}

\begin{table}[h]
\caption{Solutions associated to the fixed points of $R+\chi R^n$.
The solutions are physical only in the intervals of $p$ mentioned in
the last column.} \label{tabRRnSol}
\begin{tabular}{lccc} \hline\hline
Point & Scale Factor & Energy Density& Physical
\\ \hline
$\mathcal{A}$ & $a(t)=  \left(t-t_0\right)$ & $0$ &  $n\geq1$\\
$\mathcal{B}$ & $a(t)= a_0 \left(t-t_0\right)^{1/2}$ & $0$&  $n\geq1$\\
$\mathcal{C}$ &$a(t)= \left(t-t_0\right)$& $0$&$n\geq1$ \\
$\mathcal{D}$ &$a(t)= a_0 \left(t-t_0\right)^{1/2}$& $0$&  $n\geq1$\\
$\mathcal{E}^{*}$ & $\begin{array}{c}\left\{
                      \begin{array}{l}
                        a(t)= a_0,  \\
                        a(t)=a_0\exp\left[\pm
                        2\sqrt{3}\chi^{\gamma}(2-3n)^{\gamma}(t-t_{0})\right],\quad
                        \end{array}\right.\\\gamma=\frac{1}{2(1-n)} \end{array}$&0&$\begin{array}{c}
                                     n\geq0\\
                                    \begin{array}{c}
                                      n<\frac{2}{3}, \chi>0 \;\; \vee\\
                                      n>\frac{2}{3}, \chi<0
                                    \end{array}
                                   \end{array}$\\
$\mathcal{F}$ & $a(t)= \left(t-t_0\right)$& $0$&$n\geq1$ \\
$\mathcal{G}$ & $a(t)= a_0 \left(t-t_0\right)^{1/2}$ & 0 &  $n\geq1$\\
$\mathcal{H}$ & $a(t)=  \sqrt{1-2n(n-1)}\left(t-t_0\right)$ & $0$&$1\leq n\geq \frac{1}{2}+\frac{\sqrt{3}}{2}$\\
$\mathcal{I}^{*}$ & $a(t)=a_0\left(t-t_0\right)^{\frac{2 n^2-3
n+1}{2-n}}$  & $\mu_m=\mu_{m\,0}t^{-\frac{3 \left(2 n^2-3 n+1\right) (w +1)}{n-2}}$& $n=\frac{1}{2}, \mu_{m\,,0}=0$\\
$\mathcal{\mathcal{L}}$& $a(t)= a_0\left(t-t_0\right)^{\frac{2 n}{3
(w
+1)}}$  & $\mu_m=\mu_{m\,,0} (t-t_{0})^{2 p} $ & non physical\\
\hline\hline
 \end{tabular}
\end{table}

\section{Conclusions}
In this paper we have presented a general formalism that allows
one to apply DSA to a generic fourth order Lagrangian. The crucial
point of this method is to express the two characteristic
functions \cite{amendola:dynsys06}:
\begin{equation}
\displaystyle\Upsilon=\frac{f'}{R f'' }\;,~~~
r\,=\,-\displaystyle\frac{R f'}{f}\,,
\end{equation}
in terms of the dynamical
variables, which, in principle, allows one to obtain a closed
autonomous system for any Lagrangian density $f(R)$.

The resulting general system admits many interesting features, but
is very difficult to analyze without specifying
the function $\Upsilon$ (i.e. the form of $f(R)$).
Consequently, a ``one parameter"  approach can lead to a number
of misleading results.

Even after substituting for $\Upsilon$, the dynamical
system analysis is still very delicate; in fact, $\Upsilon$
could be discontinuous, admit singularities or generate
additional invariant submanifolds that influence deeply the
stability of the fixed points as well as the global evolution of the
orbits.

After describing the method, we applied it to two
 classes of fourth order gravity models: $R+\chi R^{n}$
and $R^{p}\exp(q R)$, finding some very interesting preliminary
results for the finite phase space. Both these models have fixed points with corresponding
solutions that admit accelerated expansion and,  consequently can model either inflation or dark energy eras
(or both). In addition, there are other fixed points which are
linked to phases of decelerated expansion which can in principle
allow for structure formation. These latter solutions are not physical for
every value of their parameters, but this is not necessarily a
problem. In fact, in order to obtain a Friedmann cosmology evolving
towards a dark energy era, these points are required to be unstable
i.e., cosmic histories coast past them for a period which depends on the
initial conditions. This means that the general integral of the cosmological
equations corresponding to such an orbit will only approximate the fixed point
solution and this approximate behavior might still allow structures to form.

It is also important to mention the fact that even if one has the
desired fixed points and desired stability, this does not necessarily imply
that there is an orbit connecting them. This is due to the presence of
singular and invariant submanifolds that effectively divide the phase
space into independent sectors. Of course one can implement further
constraints on the parameters in order to have all the interesting
points in a single connected sector, but this is still not
sufficient to guarantee that an orbit would connect them. The
situation is made worse by the fact that, since the phase space is of
dimension higher than three, chaotic behavior can also occur. It is
clear then, that any statement on the global behavior of the orbits
is only reliable if an accurate numerical analysis is performed. However,
these issues (and others) will be investigated in more detail
in a series of forthcoming papers.

A final comment is needed regarding the differences between our
results and the ones given in \cite{amendola:dynsys06}. Even if the
introduction of $\Upsilon$ and $r$, was suggested for the first
time in that paper,  the results above (and in particular the
existence of a viable matter era) are in disagreement with the ones
given in that paper. The reason is that the authors of
\cite{amendola:dynsys06} used ``a one parameter description" in
order to deal with \rf{dynsysNoK} in general. We were able to prove
that, unfortunately, not only are the equations given in
\cite{amendola:dynsys06} incomplete, but also that the method
also gives both incorrect and misleading conclusions.

\end{document}